\documentclass[conference]{IEEEtran}
\usepackage{amsmath,amssymb,amsfonts}
\usepackage{algorithmic}
\usepackage{graphicx}
\usepackage{textcomp}
\usepackage{xcolor}

\usepackage{biblatex} 
\usepackage{orcidlink}

\begin{document}

\title{GNSS SpAmming: a spoofing-based GNSS denial-of-service attack}

\author{\IEEEauthorblockN{1\textsuperscript{st} Sergio Angulo Cosín}
\IEEEauthorblockA{\textit{Space Security Centre} \\
\textit{National Institute for Aerospace Technology (INTA)}\\
Torrejón de Ardoz, Spain\\
0009-0002-8751-3361}
\and
\IEEEauthorblockN{2\textsuperscript{nd} Javier Junquera-Sánchez}
\IEEEauthorblockA{\textit{Space Security Centre} \\
\textit{National Institute for Aerospace Technology (INTA)}\\
Torrejón de Ardoz, Spain\\
0000-0002-4597-6539}
\and
\IEEEauthorblockN{3\textsuperscript{rd} Carlos Hernando-Ramiro}
\IEEEauthorblockA{\textit{Space Security Centre} \\
\textit{National Institute for Aerospace Technology (INTA)}\\
Torrejón de Ardoz, Spain\\
0000-0003-1798-1303}
\and
\IEEEauthorblockN{4\textsuperscript{th} José-Antonio Gómez-Sánchez}
\IEEEauthorblockA{\textit{Space Security Centre} \\
\textit{National Institute for Aerospace Technology (INTA)}\\
Torrejón de Ardoz, Spain\\
0009-0009-2351-7823}}

\maketitle

\begin{abstract}
GNSSs are vulnerable to attacks of two kinds: jamming (i.e. denying access to the signal) and spoofing (i.e. impersonating a legitimate satellite). These attacks have been extensively studied, and we have a myriad of countermeasures to mitigate them. In this paper we expose a new type of attack: SpAmming, which combines both approaches to achieve the same effects in a more subtle way. 

Exploiting the CDMA multiplexing present in most GNSSs, and through a spoofing attack, this approach leads the receiver to lose access to the signal of a legitimate satellite, which would be equivalent to a denial of service; but in this case the existing countermeasures against jamming or spoofing would not allow safeguarding its effectiveness, as it is neither of them.

An experimental proof-of-concept is presented in which its impact is evaluated as a function of the previous state of the receiver. Using an SDR-based system developed at the Space Security Centre, the attack is executed against a cold-started receiver, a warm-started receiver, and a receiver that has already acquired the PVT solution and is navigating. Different attack configurations are also tested, starting from a raw emission of the false signal, to surgical Doppler effect configuration, code offset, etc.
Although it is shown to be particularly successful against cold-started receivers, the results show that it is also effective in other scenarios, especially if accompanied by other attacks.
We will conclude the article by outlining possible countermeasures to detect and, eventually, counteract it; and possible avenues of research to better understand its impact, especially for authenticated services such as OSNMA, and to characterize it in order to improve the response to similar attacks.
\end{abstract}


\section{Introduction}

Global Navigation Satellite Systems (GNSS), such as GPS, Galileo, GLONASS, or BeiDou, provide essential navigation services for a multitude of applications, through constellations of satellites that transmit navigation data, primarily timestamps \cite{langley_introduction_2017}. Receivers use this data to calculate their position, velocity, and time (PVT solution), but the growing dependence on these systems has made them more attractive to attackers.

The most common attacks are jamming and spoofing. Jamming aims to produce a denial of service by generating interference that prevents the acquisition of the satellite signal in the receiving equipment. Spoofing, on the other hand, through imitation of the signal transmitted by a legitimate satellite, allows fraudulent navigation data to be sent to the receiver, causing it to calculate an erroneous PVT solution.

There are numerous countermeasures against these attacks: from characterizing the legitimate signal in order to distinguish it from interference, to authenticating the navigation message (e.g. Galileo's OSNMA service). However, these countermeasures have limitations that may allow attackers to circumvent them under certain conditions.

This paper will analyse these weaknesses and will present the SpAmming attack: a new offensive strategy that, through spoofing, manages to generate a denial of service identical to that caused by a jamming attack, but much more difficult to detect. A study of measures to mitigate its viability will also be carried out.

The study, eminently experimental, will focus on the Galileo system, mainly due to the distinctive value of the OSNMA authenticated service \cite{european_union_agency_for_the_space_programme_galileo_2023}, and the impact that its degradation may have on other GNSSs, but it is applicable to any radio system using code-based multiplexing (CDMA).

\section{Context}

\subsection{GNSS services}

GNSSs have the main objective of allowing the receiver to resolve its position, velocity and time (PVT). For this purpose, the radio signal is processed as follows \cite{borre_software-defined_2007}:

\begin{itemize}
    \item Acquisition: the transmitting satellites are identified by their PRN and the parameters involved in the signal (phase shift and Doppler effect) are ascertained. In this paper, we will work assuming the use of the PCPS acquisition method \cite{romani_exploring_2022}, consisting of a first search, with low resolution, of the delay and Doppler; followed by a second, more refined search, based on the data obtained by the first one.
    \item Tracking: once the emitting satellites and the amount of delay and Doppler effect affecting the signal have been identified, synchronization with the signal must be maintained, and these parameters must be updated dynamically \cite{lezaeta_collazos_software_2010}.
    \item Decoding: demodulation of the signal, obtaining the navigation message bits. After decoding, the observables are determined, i.e. the key parameters to calculate the PVT: pseudorange, code and carrier offset, and Doppler effect.
    \item Calculation of the PVT solution: using the data obtained from the GNSS signal and available satellite orbit information, the receiver calculates its position (i.e. distance calculation, as a function of pseudorange), velocity as a function of Doppler, and time, fixing the one emitted by the satellite.
\end{itemize}

\subsection{Attacks}

Jamming consists in the generation of random noise in the band to be denied and is usually aimed at preventing the establishment of a time and position reference, either to deny service or to increase the viability of other attacks (in fact, it is usually an indicator for spoofing detection). It is very inefficient and easily detectable, but COTS equipment exists both for the generation and the detection of jamming attacks.

Regarding spoofing attacks, the aim is to impersonate a legitimate satellite with different levels of complexity, depending on the assumed capabilities of the victim: from broadcasting exclusively with its PRN, to emulating the complete navigation message. Depending on the particular system, there may be open source tools for this (e.g. gps-sdr-sim \cite{ebinuma_osqzssgps-sdr-sim_2025}), and the concrete implementation of the attack may be \cite{islam_impact_2023} 

\begin{itemize}
    \item single v. multitransmitter: depending on the quantity of signal sources.
    \item synchronous v. asynchronous: depending on whether the spoofing signal is synchronised with the legitimate signal.
    \item continuous v. intermittent: depending on the duration of the disruptions.
    \item static (i.e. abrupt) v. dynamic (i.e. with gradual changes).
\end{itemize}

Finally, meaconing attacks take advantage of the relevance of time in the PVT solution process, through a replay attack. By replaying a previously recorded real signal, the same effects as with a spoofing attack would follow, in a simple but not very adaptable way.

\subsection{Countermeasures}

GNSS services can be protected against attackers in several ways: 

\begin{itemize}
    \item Prevention: Most commonly, some elements of the signal are encrypted, although other strategies have recently been adopted. The European GNSS, Galileo, has several protected services: PRS, CAS, and OSNMA \cite{gomez-sanchez_soluciones_nodate}. The latter is part of the Galileo Open Service, but has the particularity that the message is authenticated, so that the receiver can be sure that the received signal really comes from Galileo, protecting itself from possible spoofing attacks. Only some satellites in the constellation have the OSNMA service \cite{fernandez-hernandez_navigation_2016}, which makes the whole service vulnerable to SpAmming attacks, as we shall see.
    \item Detection: one of the most obvious effects of this type of attack is the abrupt drop in the signal-to-noise ratio (C/N0) \cite{do_prado_allao_systematic_2025}, or the comparison between the network time and the time emitted by the GNSS system, in equipment connected to the Internet.
    \item Response and recovery: the most effective measure to mitigate this type of attacks is by using antenna systems with which directional screening can be performed, allowing the elimination of malicious components of the signal, such as with CRPA antennas \cite{vegni_gnss_2015}.
\end{itemize}

\section{Materials and methods}

\subsection{Test environment}

To evaluate the effectiveness of a SpAmming attack, a test environment has been prepared with the following elements:

\begin{itemize}
    \item Leica AR20 antenna installed on the roof of the building.
    \item SDR platform: Ettus USRP B210, which functions as the SpAmming signal emitter.
    \item Connector: Power splitter minicircuits ZAPD-2-S+, used to sum the SpAmming signal and the aerial signal coming from the antenna.
    \item DC blocking filter.
    \item GNSS receiver: u-blox ZED-F9P, on u-blox C099-F9P development board. 
\end{itemize}

The software used is:

\begin{itemize}
    \item gal-sdr-sim: Python 3 project, developed by the Space Security Centre to create the baseband signals.
    \item GNU Radio Companion: to transmit the signals in modulated band.
    \item u-blox u-center: to analyse the received GNSS signals
\end{itemize}

\begin{figure}[ht]
    \centerline{\includegraphics[width=0.45\textwidth]{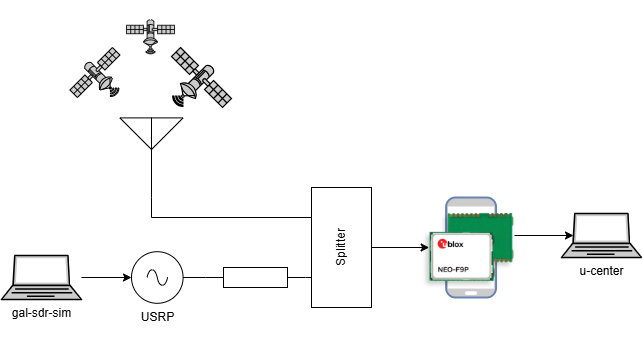}}
    \caption{Experiment components design.}
    \label{fig:design}
\end{figure}

In all the tests the signals have been transmitted by cable, following the scheme shown in Figure \ref{fig:design}, to avoid creating unintentional interference that could affect other users.

\begin{figure*}[htbp]
    \centerline{\includegraphics[width=0.9\textwidth]{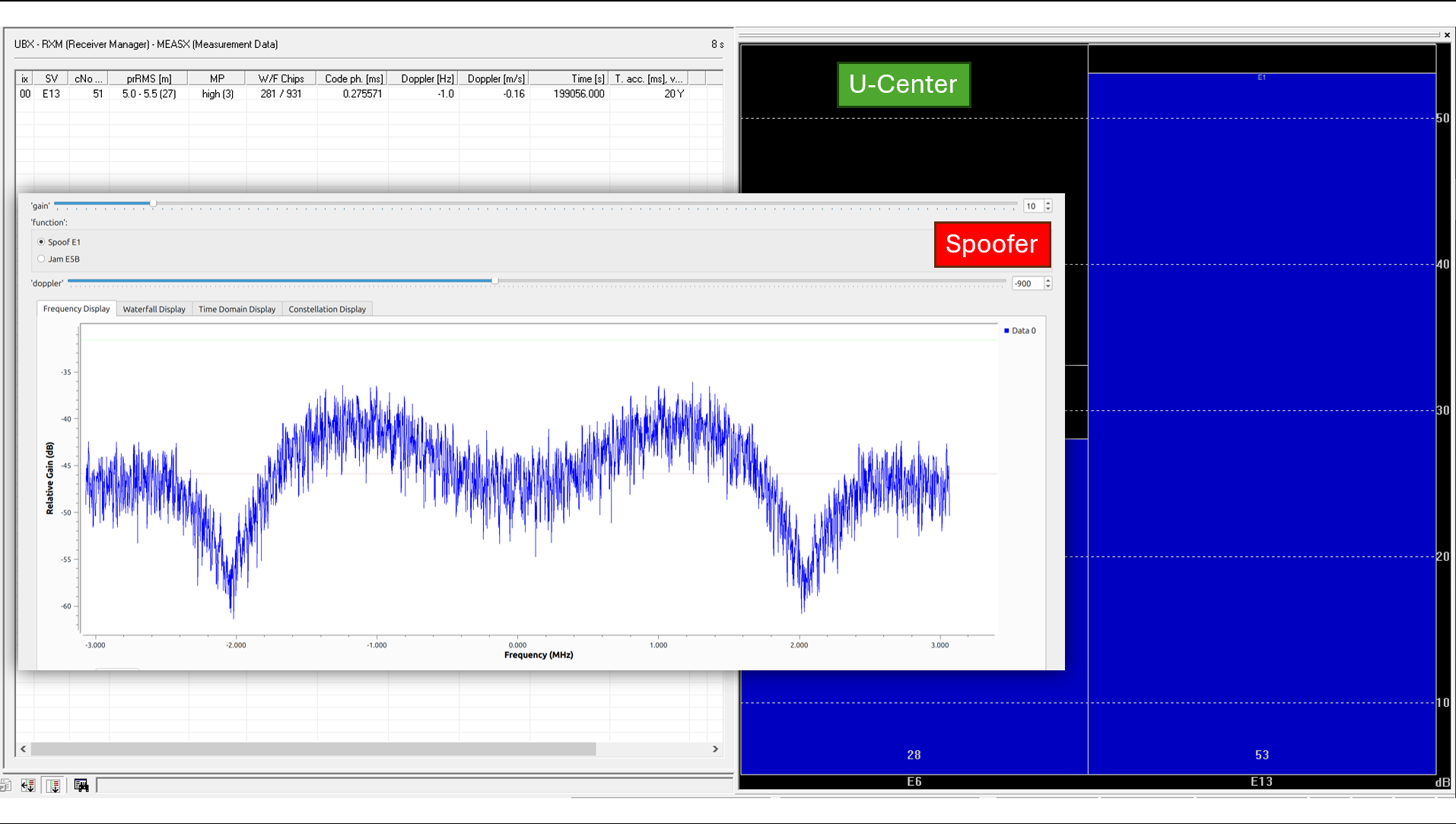}}
    \caption{Atacker interface v. resulting efects.}
    \label{fig:spoof-n-res}
\end{figure*}

The complete attacker interface can be shown in Figure \ref{fig:spoof-n-res}, over the results displayed by the u-blox data visualization tool (i.e. u-center).

\subsection{SpAmming strategy}

A SpAmming attack consists of performing a denial of service through spoofing, instead of introducing noise into the signal. By spoofing only a few satellites, and making their acquisition impossible or causing a disruption in tracking, the receiver can be forced to use the remaining satellites to obtain a PVT solution. 

To succeed in interfering with a particular satellite, without affecting the reception of the signal coming from the rest, some characteristics of the genuine satellite signal will be mimicked, such as the modulation and multiplexing parameters (i.e. the particular PRN of the satellite to be supplanted). However, although desirable in practice, it is not necessary to mimic the entire signal in particular: there is no requirement to create a coherent navigation message, instead it would be sufficient to perform the spoofing at the CDMA level. 

For these reasons, it is hypothesized that SpAmming is simpler to carry out than a spoofing attack, and is less detectable than a traditional jamming attack, although it results in the same situation: a denial of service.

\subsection{Tested scenarios}

Several attack scenarios were tested, with different SpAmming signal configurations, and taking into account the particular state of the receiver with respect to the signal in space:

\begin{itemize}
    \item Cold start: start the receiver without any information, and try to attack the signal acquisition process, for which it should suffice to emit a false signal from any of the satellites in view at that time.
    \item Warm start: when the receiver is started, it knows the time, although not very accurately, and has the almanac. It will be necessary to substitute the satellites that are visible at the time of execution, but taking into account that the receiver may have enough information to infer the Doppler and the code offset.
    \item Hot start: the receiver is in full operation, or at the start up it has the precise time, almanac, and updated ephemeris. In this scenario we will attack the signal tracking process, trying to supplant the acquisition parameters (e.g. Doppler, code offset, etc.), plus the carrier offset, working at the chip level.
\end{itemize}

Our attack was launched on the signal of a particular satellite, evaluating if the receiver: \begin{enumerate}
    \item loses access to the signal, 
    \item is able to recover it for the duration of our attack, and, 
    \item is able to recover it when our attack ends.
\end{enumerate}

\section{Results}

For each of the scenarios and strategies proposed, we applied the following procedure:

\begin{enumerate}

    \item Once the receiver was connected to the genuine signal, we could identify which satellites were in view. We then selected one of these satellites and made a note of:

    \begin{enumerate}
        \item Vehicle identifier (SVID), i.e. satellite number in the constellation.
        \item Code offset with respect to the receiver, in ms.
        \item Doppler detected in the receiver, in Hz.
    \end{enumerate}

	\item We generated our own signal with gal-sdr-sim, choosing as the target SVID the one selected in the previous step.
    
    \item Using the splitter, we mixed our signal with the nominal signal and dynamically configured the necessary parameters (e.g. Doppler, power, offset, etc.) to adjust the transmission to the specific evaluation we are performing.

\end{enumerate}

\begin{figure}[htbp]
    \centerline{\includegraphics[width=0.45\textwidth]{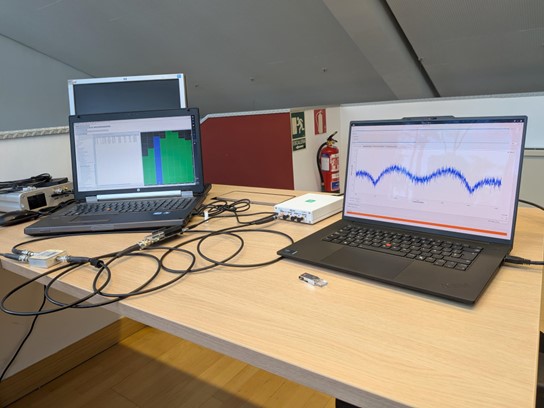}}
    \caption{Laboratory setup for evaluating the SpAmming attack}
    \label{fig:setup}
\end{figure}

Figure \ref{fig:setup} shows the configuration of the environment in which the tests were carried out. On the left there is the equipment used to evaluate the results with u-center. On the right there is the equipment used to perform the spoofing. In between there are the radio devices (i.e. u-blox, USRP, etc.).

\subsection{Experiment 1: cold start}

\begin{figure*}[htbp]
    \centerline{\includegraphics[width=0.9\textwidth]{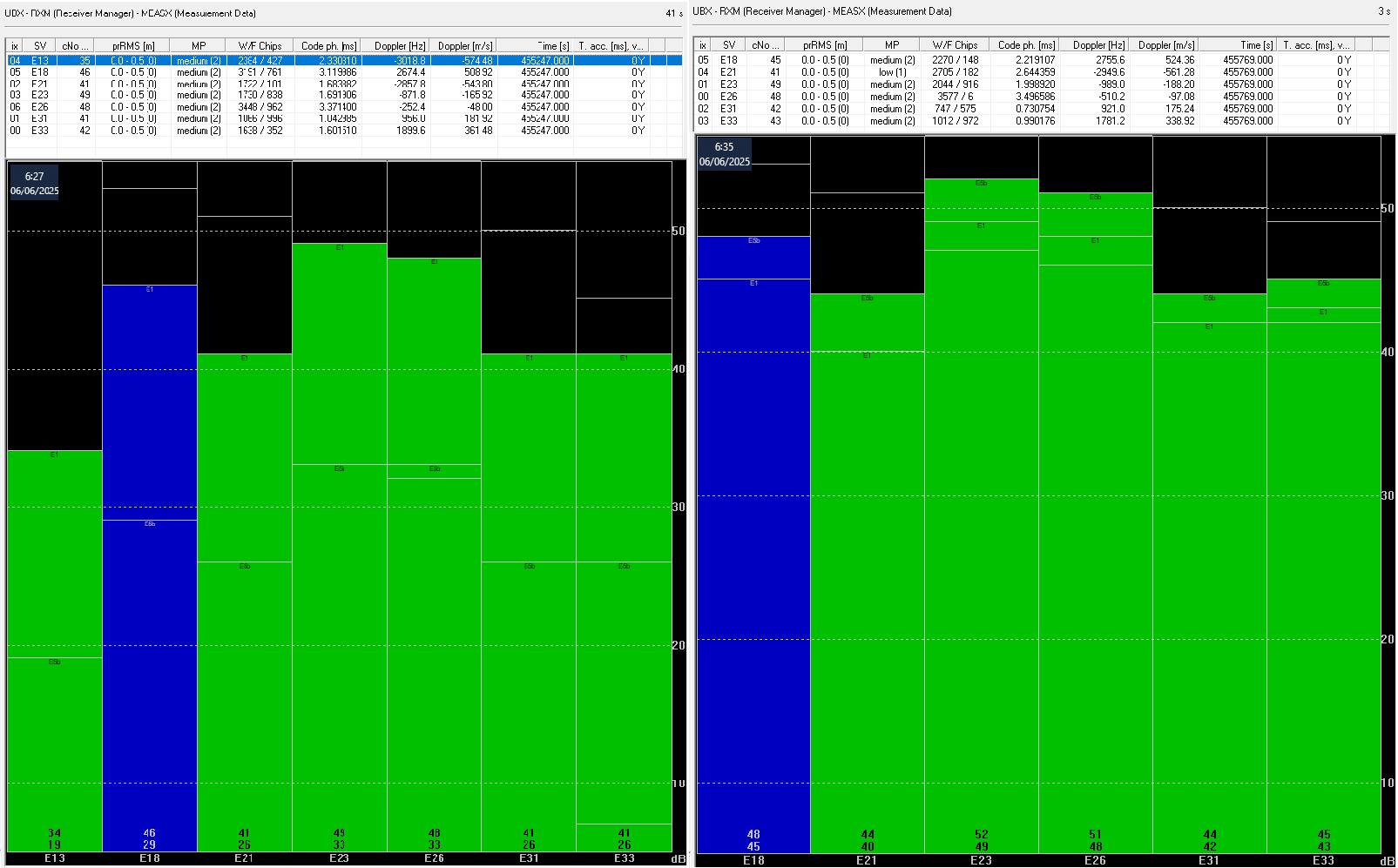}}
    \caption{Nominal scenario v. Scenario after cold start and SpAmming}
    \label{fig:cold-result}
\end{figure*}

We started the receiver in a cold state and launched our attack against the satellite with SVID 13. Since we did not enter a real Doppler value, when we stopped the attack, the receiver only searched for the satellite with the false Doppler value and could not find it. Figure \ref{fig:cold-result} shows the nominal results compared with those obtained after the attack.
 
\subsection{Test 2: warm start}

To execute this scenario, we analysed the legitimate signal and transmitted the signal from the supplanted satellite with the same Doppler shift. The USRP introduces a base Doppler shift of around $1.2$ kHz, so, since even a small change could cause the attack to fail, this had to be taken into account when generating the signal.

Furthermore, since the E5B signal has characteristics similar to the legitimate signal, the receiver acquired it in parallel. Therefore, jamming was included in that band to limit the test scenario to E1.

\subsection{Test 3: hot start}

This test evaluated an attack on a switched-on receiver. To achieve this, we obtained and reproduced the parameters of the signal reaching the receiver (i.e. Doppler shift, phase shift, etc.). Then we adjusted the transmission to them as much as possible.

The attack has occasionally been successful. However, if intermittent jamming was also performed in the E1 band during exercise, our effectiveness improved considerably, achieving effects similar to those obtained in the warm start execution.

\section{Discussion}
From a cold start, our attack is practically guaranteed to succeed and have permanent effects on the receiver. If we focus on satellites offering protected GNSS services, such as OSNMA, the receiver will be forced to use only unauthenticated satellites. In this way, a standard spoofing attack could easily be performed.

Unlike with a cold start, with a warm start we must maintain the attack for as long as we want to deny the satellite access. When we stop the attack, the satellite reacquires the legitimate signal. Furthermore, SpAmming would have to be accompanied by jamming the E5B band, or preferably supplanting the signal in that band, also.

Finally, the hot-start scenario is the most resilient. Although emulating the Doppler shift of the legitimate signal is relatively simple, implementing code offset is somewhat more difficult. Given the tools used (i.e. software-defined radio), adjusting the carrier offset would be almost impossible. Further experimentation would be necessary to avoid revealing our attack through interference, perhaps using different methods.

\subsection{Countermeasures}

In conclusion, our approach has identified two weaknesses in current GNSSs. Here we propose some countermeasures.

Firstly, an attacker could disable the OSNMA entirely by impersonating the satellites that emit it. This could be addressed by implementing authentication in the entire system, instead of using a subset of satellites.

Secondly, message authentication alone is insufficient. The most effective countermeasure would be authentication at the PRN level, as proposed in the GPS Chimera service \cite{nicola_gps_2021}. SpAmming could easily be detected in this manner.

\section{Conclusion and future work}

This article presents a new method of attacking GNSS signals to deny service to specific satellites through spoofing. This technique is known as SpAmming. The effectiveness of this method has been evaluated in various scenarios, and countermeasures to mitigate its effects have been presented. These countermeasures mainly involve authenticating the signal at the PRN level rather than just the navigation message.

\subsection{Future work}

As this is solely a proof of concept, a formal evaluation is a task for future work. In view of the limitations of the technique when not starting from a cold start, it would be useful to ascertain whether the same permanent effects could be achieved using a dynamic approach. Finally, evaluating the impact on other systems of using SpAmming against OSNMA satellites would demonstrate its relevance, as it would pave the way for a traditional spoofing attack.

\printbibliography

\end{document}